%% file: main.tex
\documentclass[conference,hidelinks]{IEEEtran}
\IEEEoverridecommandlockouts

\input{load-packages}

\newcommand*{\RT}{RT}

\newcommand*{\Nfft}{N_{\text{stft}}}
\newcommand{\posterior}[2]{\mathcal{N}\left( #1 \mid \mu_{#2}, \Sigma_{#2} \right)}

\newcommand{\lnorm}[1]{ \left\lVert #1 \right\rVert }
\newcommand{\abs}[1]{ \left\lvert #1 \right\rvert }

\newacronym{ai}{AI}{Artificial Intelligence}
\newacronym{bic}{BIC}{Bayesian information criterion}
\newacronym{dsp}{DSP}{Digital Signal Processing}
\newacronym{gmm}{GMM}{Gaussian Mixture Model}
\newacronym{mfcc}{MFCC}{Mel Frequency Cepstral Coefficient}
\newacronym{rir}{RIR}{Room Impulse Response}
\newacronym[plural={$\RT_{60}$ values}]{rt60}{$\RT_{60}$}{reverberation time}
\newacronym{svm}{SVM}{Support Vector Machine}
\newacronym{rbf}{RBF}{Radial Basis Function}

\newcommand{\splitcell}[2][c]{%
  \begin{tabular}{@{}#1@{}}
  #2
  \end{tabular}%
}

\makeatletter
\newcommand{\warn}[1]{\@latex@warning{#1}}
\makeatother

\begin{document}

\input{front-matter/title-and-authors}
\maketitle
\input{front-matter/abstract-and-keywords}

\input{sec-intro}
\input{sec-background}

\section{Proposed Approach}\label{sec:approach}
\subsection{Overview}
Our approach for environment classification consists of four main steps, summarised here and described in the following sections:
\begin{enumerate}
    \item A \emph{blind channel magnitude estimation} step, which estimates the magnitude $\lvert{\widehat H(f)}\rvert$ of the \gls{rir} associated to the recording $x(t)$
    \item A \emph{minimum-phase digital filter estimation} step, which determines a digital filter $\widehat H(z)$ with minimum-phase response and magnitude as close as possible to $\lvert{\widehat H(f)}\rvert$
    \item A \emph{blind roomprint estimation} step, which computes the roomprint $\hat \phi$ associated to output in the time domain of the digital filter $\widehat H(z)$
    \item A \emph{closed-set classification} step, which according to the the estimated roomprint $\hat \phi$ returns a \emph{label} of the room in which the recording $x(t)$ took place.
\end{enumerate}

\input{sec-approach-channel}
\input{sec-approach-minimum-phase}
\input{sec-approach-roomprint}
\input{sec-approach-svm}
\input{sec-evaluation}
\input{sec-outlook}

\warn{TODO: Balance last page for camera-ready version}
\printbibliography


\end{document}

%% file: load-packages.tex
\usepackage{amsmath,amssymb,amsfonts}
\usepackage{algorithmic}
\usepackage{graphicx}
\usepackage{textcomp}
\usepackage{xcolor}
\def\BibTeX{{\rm B\kern-.05em{\sc i\kern-.025em b}\kern-.08em
    T\kern-.1667em\lower.7ex\hbox{E}\kern-.125emX}}

\usepackage[style=ieee,mincitenames=1,maxcitenames=2]{biblatex}
\AtBeginBibliography{\footnotesize}

\usepackage{hyperref}
\usepackage{cleveref} 
\usepackage{glossaries}
\usepackage{booktabs, multirow, multicol}
\usepackage{graphicx}
\usepackage{subcaption}
\glsdisablehyper

\usepackage{orcidlink}

\usepackage{nicefrac}

%% file: front-matter/title-and-authors.tex
\title{Environment Classification \\via Blind Roomprints Estimation\\
\thanks{This paper was supported by the BMBF SpeechTrust+ project (grant no 13N16267) %
{\textcolor{black}{and by the EU H2020 AI4Media project (grant no 951911)}}%
.}
}

\author{
\IEEEauthorblockN{%
  Malte Baum\IEEEauthorrefmark{1}\IEEEauthorrefmark{2}, %
  Luca Cuccovillo\IEEEauthorrefmark{1}\,\orcidlink{0000-0001-5559-6508}, %
  Artem Yaroshchuk\IEEEauthorrefmark{1} and 
  Patrick Aichroth\IEEEauthorrefmark{1}\,\orcidlink{0000-0003-4777-6335}%
}
\IEEEauthorblockA{}
\IEEEauthorblockA{\IEEEauthorrefmark{1}\textit{Fraunhofer Institute for Digital Media Technology}, 98693 Ilmenau, Germany}
\IEEEauthorblockA{\IEEEauthorrefmark{2}\textit{Ilmenau University of Technology}, 98693 Ilmenau, Germany}
}

%% file: front-matter/abstract-and-keywords.tex
\begin{abstract}
In this paper we present a novel approach for environment classification for speech recordings, which does not require the selection of decaying reverberation tails. It is based on a multi-band RT60 analysis of blind channel estimates and achieves an accuracy of up to 93.6\% on test recordings derived from the ACE corpus.
\end{abstract}

\begin{IEEEkeywords}
environment classification, audio forensics, multi-band RT60 estimation
\end{IEEEkeywords}

%% file: sec-intro.tex
\section{Introduction}

Much of modern communication takes place via social networks, in real-time, and with the help of smartphones and other mobile devices~\cite{PewResearch:2019}, resulting in an ever-increasing amount of user-generated content, and potential misuse: For instance, recordings can be manipulated in order to spread rumors and fake news, to support phishing attacks, or to create forged evidences to influence a trial. One discipline that can support content verification and detect such manipulations is audio forensics, which aims at analyzing footprints left by the recordings process, using automatic and reproducible analysis algorithms.

Environment classification is a specific audio forensics technique, which aims at determining where a recording took place -- e.g., outdoor, in a small room, in a large church, by exploiting the acoustic characteristics of the recording environment~\cite{Roomprints:AES:2014,Roomprints:IEEE:2013}. Such information can be very useful for verification, e.g. in court cases, as it supports the comparison of \textit{claims} against \textit{facts} regarding environment and recording characteristics. However, this task is challenging, and proposed approaches have struggled to reach an acceptable level of accuracy for handling real-life speech material~\cite{EnvClass:Patole:CAST,EnvClass:Malik:ICASSP}. 

Furthermore, while some research focused on detecting inconsistencies of acoustic parameters related to decaying reverberation tails of a recording, thereby identifying spliced speech segments within a pre-existing recording~\cite{Capoferri:2020:EnvSplicingDetection,Patole:2017:EnvSplicingDetection}, no attention has yet been devoted to environment classification in presence of speech signals: Existing literature requires the presence of impulsive signals created e.g. by hand claps to work, which however rarely occur in relevant audio recordings~\cite{Roomprints:AES:2014,Roomprints:IEEE:2013,EnvClass:Patole:CAST,EnvClass:Malik:ICASSP}.

In this paper, we propose to address the problem by applying multi-band \gls{rt60} analysis of blind channel estimates: Multi-band \gls{rt60} analysis has been proposed as essential component for deriving roomprint features for environment classification in~\cite{Roomprints:AES:2014,Roomprints:IEEE:2013}. Within the original proposed approach, such features were however retrieved under laboratory conditions, via analysis of the noiseless room impulse response in the time domain. Our proposal, in contrast, uses blind estimate of the existing transmission channel~\cite{Gaubitch:ChannelEstimation}, which is used as source for estimating the desired roomprints. As a consequence, it is applicable to speech recordings without the need for automatic or semi-automatic selection of decaying reverberation tails, which is an error-prone step potentially leading to increased error rates.

The paper is structured as follows: In \Cref{sec:background}, we present the background terminology and concepts required to retrieve roomprints for environment classification, which we then use in the algorithm proposed in \Cref{sec:approach} to perform environment classification from speech recordings, rather than in laboratory conditions. In \Cref{sec:evaluation}, we then evaluate the classification algorithm using recordings derived from the ACE corpus~\cite{dataset:ace-challenge}, then closing in \Cref{sec:outlook} with a summary and possible future research steps.

%% file: sec-background.tex
\section{Background}\label{sec:background}

\subsection{Room Impulse Response and Reverberation Time}

Let us denote with $s(t)$ an input speech signal reverberating through an environment characterized by a \gls{rir} denoted by $h(t)$. The corresponding signal $x(t)$ recorded by the receiving microphone can be modeled by means of a convolution $*$ in the time domain: 
\begin{equation}\label{eq:rir-model}
    x(t) = h(t) \ast s(t) + n_\text{env}(t),
\end{equation}
where $n_\text{env}(t)$ denotes the environmental noise and is therefore assumed equal to zero for noiseless recordings. 

\begin{figure}[ht]
  \centering
  \includegraphics[width=0.8\columnwidth]{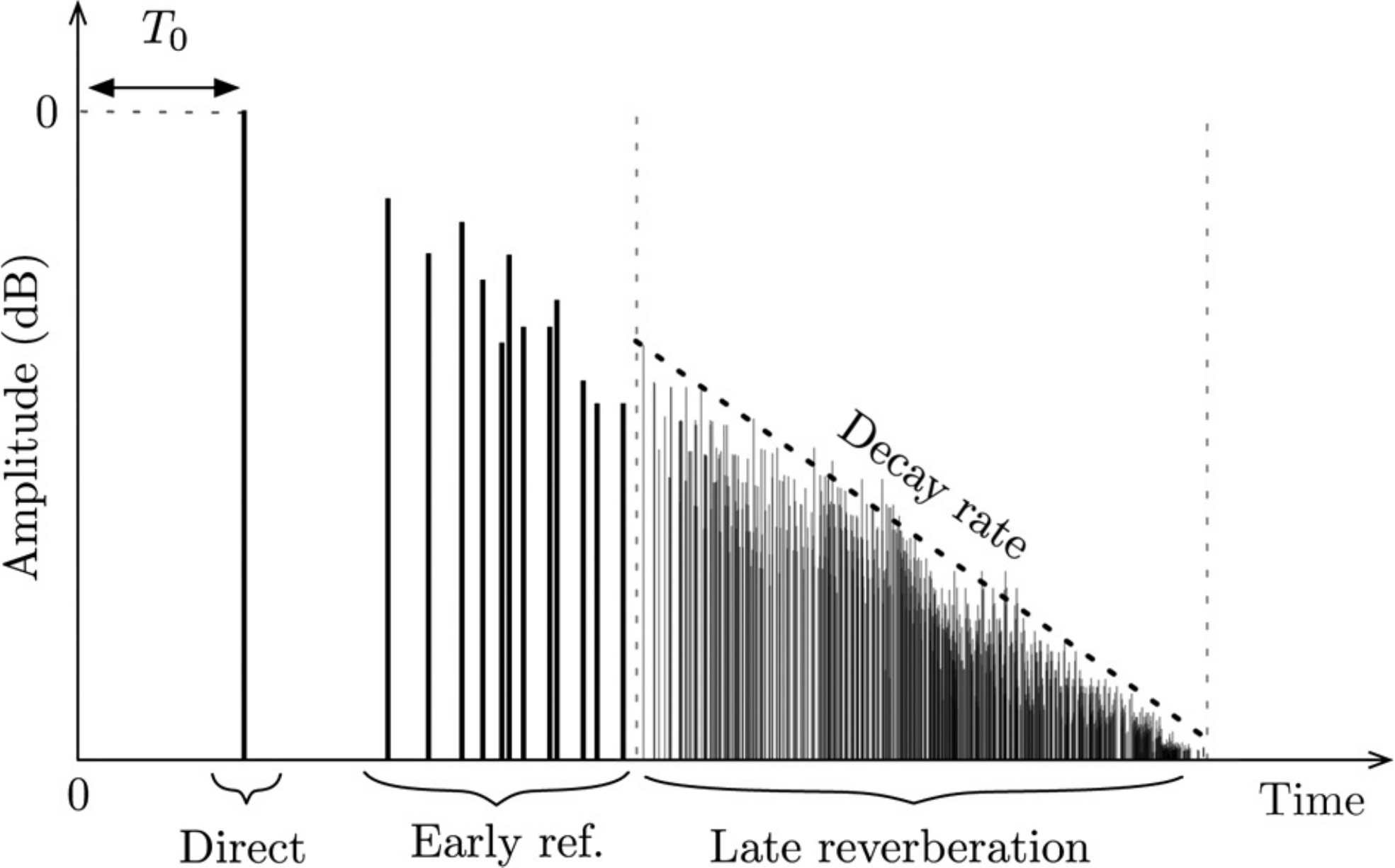}
  \caption{Schematic example of a generic room impulse response in which $s(t)$ is a unitary pulse~\cite{Valimaki:2012:RoomImpulseResponse}}
  \label{fig:rir}
\end{figure}

As depicted in \Cref{fig:rir}, the \gls{rir} $h(t)$ not only defines the time-delay $T_0$ between the sound emission and its acquisition, but most importantly the amount and characteristics of \emph{early reflections}, conveying most of the information on the geometry and materials of the surroundings, and of \emph{late reverberations}, conveying information on the size of the surroundings and of the absorbing power of the materials within it.

An important parameter related to the \gls{rir} is the reverberation time $\RT_{60}$, i.e., the amount of time required for the space-averaged sound energy density in an
enclosure to decrease by 60 dB after the source emission has stopped~\cite{ISO:RT60}:

\begin{equation}\label{eq:reverb:rt60}
\RT_{60}=\left(t~~:~~10\log_{10}\left(\left|\frac{x(0)}{x(t)}\right|\right)=60\right).
\end{equation}

In practical cases, due to the presence of noise, $\RT_{60}$ is derived by linear interpolation:
\begin{equation}\label{eq:reverb:rt20}
\RT_{60}=\alpha\cdot \RT_{60 / \alpha},
\end{equation}
where the parameter $\alpha$ can be adapted according to the noise conditions. Estimation is therefore often performed according to $\RT_{40}$ ($\alpha=$ \nicefrac{3}{2}), $\RT_{30}$ ($\alpha=2$) or  $\RT_{20}$ ($\alpha=3$).

\subsection{Roomprints for Environment Classification}

Roomprints for environment classification, introduced by \citeauthor{Roomprints:IEEE:2013} in \cite{Roomprints:IEEE:2013} and then refined in \cite{Roomprints:AES:2014}, propose to characterise a room by means of multi-band \gls{rt60} estimation. 

The authors argued that the \gls{rt60} is largely related to the absorption properties of the materials present in the acoustic environment, which is frequency-dependent. Therefore, they proposed not to use a full-band \gls{rt60} estimation to characterize an acoustic environment, but rather to do so according to fractional octave filterbanks.

Fractional octave filterbanks can be calculated according to the ANSI standard \cite{ANSI:OctaveBand} with the base-two system. Let us denote with $B$ a positive integer to designate the fraction of an octave band (\nicefrac{1}{B}). 

The exact midband frequency of each bandpass filter of the filterbank is given by
\begin{equation}
f_m^{(b)} = 
\begin{cases}
  2^{(b-30)/B} f_r     & \text{if $B$ is odd,} \\
  2^{(2b-59)/(2B)} f_r & \text{if $B$ is even.}
\end{cases}
\end{equation}
$f_r$ is the reference frequency of 1\,kHz and $b$ is any integer positive, negative or zero indicating the band number. 

The frequencies of the lower and upper edges of the passband of the bandpass filter, called bandedge frequencies, can be expressed as
\begin{equation}
  f_l^{(b)} = 2^{-1/(2B)} f_m^{(b)}
\end{equation}
for the lower bandedge frequency and
\begin{equation}
  f_u^{(b)} = 2^{1/(2B)} f_m^{(b)}
\end{equation}
for the upper bandedge frequency. 

The exact midband frequency is the geometric mean  of the lower and upper bandedge frequencies:
\begin{equation}
  f_m^{(b)} = \sqrt{f_l^{(b)} \cdot f_u^{(b)}}.
\end{equation}

In order to estimate a roomprint, \citeauthor{Roomprints:IEEE:2013} proposed to first apply the $b$-th filterbank to the \gls{rir} $h(t)$, and thus obtaining the fractional octave response $h^{(b)}(t)$ of the $b$-th sub-band:
\begin{equation}
    h^{(b)}(t) = \operatorname{bandpass-filter}\left( h(t),
    f_l^{(b)}, f_u^{(b)}\right).
\end{equation}
Then, they proposed to compute the reverberation time independently for each band, obtaining one measurement $\RT_{60}^{(b)}$ per each band.

\begin{figure}[ht]
    \centering
    \includegraphics[clip,trim={2.5cm 4.0cm 3.5cm 6.5cm},width=\columnwidth]{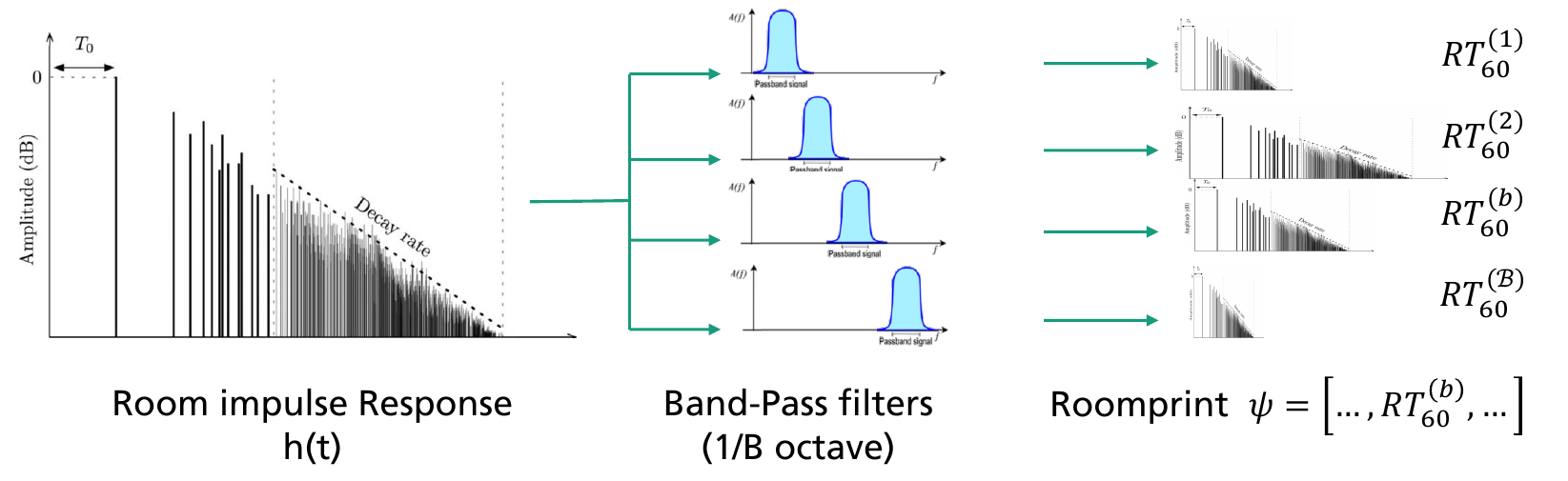}
    \caption{High level schema of roomprint extraction}
    \label{fig:roomprints}
\end{figure}

The roomprint feature $\psi$ of a single room can therefore be derived by aggregating the entire set of \gls{rt60} values: 
\begin{equation}
   \psi = \left[ \RT_{60}^{(1)}, \RT_{60}^{(2)}, \ldots, \RT_{60}^{(b)}, \ldots,  \RT_{60}^{(\mathcal{B})} \right],
\end{equation}
where each $\RT_{60}^{(b)}$ can be obtained by reverse integrating the energy of the band-passed filtered response $h^{(b)}(t)$, according to Schroeder's integration method~\cite{Schroeder:RT60}. A high-level schema of the process is depicted in \Cref{fig:roomprints}. Further details, including multi-path considerations and alternative definitions for the vector can instead be found in \cite{Roomprints:IEEE:2013,Roomprints:AES:2014}.

Roomprints have been tested extensively by the original authors, reaching a remarkable accuracy of 96.1\% for a test set of 22 rooms, using $\psi'=\log(\psi)$ as feature vector and a \gls{gmm} for classification. 

The main limitation of this feature and respective tests, however, lies in the fact that the test signals were obtained in noiseless laboratory environments by recording reverberations from sudden, impulsive bursts. The case of speech input signals was not addressed, due to the difficulties inherent its bandwidth limitation, and to the known issues related to the high variability of \gls{rt60} values estimated from decaying reverberation tails selected within speech recordings~\cite{EnvClass:Malik:ICASSP}. In the next section, we will therefore focus on the blind estimation of the room impulse response $h(t)$ from speech recordings, and retrieve a signal $\hat h(t)$ which can be used in conjunction with roomprints for effective environment classification.

%% file: sec-approach-channel.tex
\subsection{Blind Channel Magnitude Estimation}
Under the assumption of time-invariant \gls{rir}, each frame $x_l(t)$ of the input recording $x(t)$ can be modeled by re-applying \cref{eq:rir-model} to the specific analysis window:
\begin{equation}
x_l(t)=h(t)\ast s_l(t) + n_{\text{env},l}(t)
\end{equation}

An equivalent formulation in the log-power domain for the noiseless case of $n_{\text{env},l}(t)$ is:
\begin{equation}
X_l(f)=H(f) + S_l(f).
\end{equation}
If we can estimate the dry input speech $S_l(f)$, i.e., if we can compute a term $\widehat{S}_l(f)$ which is accurate enough, $h(t)$ can be estimated blindly by applying: 
\begin{equation}
\widehat{H}(f)=\frac{1}{L}\sum_{l=1}^L \left(X_l(f)-\widehat{S}_l(f)\right),\label{eq:channel-estimation:pre-final}
\end{equation}
with $L$ denoting the amount of frames and thus \cref{eq:channel-estimation:pre-final} denoting the average difference between the input recording frames and the estimated ideal speech.

To further improve the stability of \cref{eq:channel-estimation:pre-final}, both terms $X_l(f)$ and $\widehat{S}_l(f)$ may be normalized to have zero mean, by applying
\begin{subequations}
\begin{align}
   {\underbar X}_l(f) & =  X_l(f) - \frac{1}{\Nfft}\sum_{f=1}^{\Nfft} X_l(f),\\
   {\widehat{ \underbar S}_l}(f) & = \widehat{S}_l(f) - \frac{1}{\Nfft}\sum_{f=1}^{\Nfft} \widehat{S}_l(f),
\end{align}
\end{subequations}
and obtaining the final \cref{eq:channel-estimation} for the mean-normalized blind estimation of the room impulse response: 
\begin{equation}
\widehat{\underbar H}(f)=\frac{1}{L}\sum_{l=1}^L \left( {\underbar X}_l(f) - {\widehat{ \underbar S}_l}(f) \right).\quad \label{eq:channel-estimation}
\end{equation}

Intuitively, the better the estimated $\widehat{S}_l(f)$ is, the more accurate the estimated channel log-magnitude $\widehat{\underbar H}(f)$.

The basis for estimating $\widehat{S}_l(f)$ have been set by \citeauthor{Gaubitch:ChannelEstimation} in \cite{Gaubitch:ChannelEstimation}, both for the noiseless case we outlined so far, as well as for the case with additive noise. In the following, we are going to describe the estimation of the ideal dry speech in noiseless conditions and thus avoid a more cumbersome notation. If the noise is not negligible, it is possible to either follow the noise-aware estimation in \cite{Gaubitch:ChannelEstimation}, or -- according to the outcome of recent experiments on the very same channel estimation procedure applied for estimating microphone frequency responses~\cite{Cuccovillo:2022:MAD,Giganti:2022:Euspico} -- to first apply denoising in the spectral domain and then to follow the noiseless estimation procedure we are going to describe herein.

The first step of the estimation procedure consists of processing a large speech corpus to extract a high amount of RASTA filtered \glspl{mfcc}~\cite{RASTA_filter}. In the following, \glspl{mfcc} of the $l$-th frame of an input audio signal $x$ will be denoted by the symbol $c_{X_l}$. 

Given $L_X$ training \gls{mfcc} vectors $c_{X_l}$, used to fit a \gls{gmm} with $M$ mixtures, a key element of the estimation procedure is the relative mixture probability $p_i\left(c_{X_l}\right)$, i.e., the probability that the feature vector $c_{X_l}$ belongs to the $i$-th mixture:
\begin{equation}
    p_i\left(c_{X_l}\right)= \frac{
      \pi_i \cdot \posterior{c_{X_l}}{i}
    }{
      \sum_{m=1}^M \pi_m \cdot \posterior{c_{X_l}}{m}
    }.\label{eq:relative_prob}
\end{equation}
In \cref{eq:relative_prob}, $\posterior{c_{X_l}}{i}$ denotes the posterior probability of the vector $c_{X_l}$ against the $i$-th mixture, having a normal distribution with mean $\mu_i$, covariance $\Sigma_i$, and prior $\pi_i$.

With the help of these definitions, a model of the average log spectrum of the ideal speech can be obtained as follows: 
\begin{subequations}
\begin{enumerate}
    \item Build a first normalized power spectrum matrix $\underbar {X}$, by collecting row-wise all mean-normalized log powers ${\underbar X_l}(f)$ of the \gls{gmm} training set:
    \begin{equation}
        \underbar {X}\in\mathbb{R}^{L_X\times\Nfft} = \left\lbrace {\underbar X_l}(f) \right\rbrace,
    \end{equation}
    \item Build a relative probability matrix $\underbar P_{X}$, by collecting row-wise all relative mixture probabilities $p_i\left(c_{X_l}\right)$ of the \gls{gmm} training set:
    \begin{equation}
        \underbar P_{X}\in\mathbb{R}^{L_X\times M} = \left\lbrace p_i\left(c_{X_l}\right) \right\rbrace\label{eq:relative_probability_matrix}
    \end{equation}
    \item Compute the average speech spectrum matrix $\underbar S_{X}$:
    \begin{equation}
        \underbar S_{X}\in\mathbb{R}^{M\times \Nfft} = \underbar P_{X}^t \cdot \underbar X,
    \end{equation}
    with $t$ denoting the transposition.
\end{enumerate}
\end{subequations}
The matrix $\underbar S_{X}$ is at the core of the ideal speech estimation procedure in \cite{Gaubitch:ChannelEstimation}: Given an arbitrary input speech signal $s$ having $L_S$ frames and a relative probability matrix $\underbar P_{S}$, it is straightforward to compute: 
\begin{equation}
    {\widehat{\underbar S}} \in \mathbb{R}^{L_S\times\Nfft} = \underbar P_{S} \cdot \underbar S_{X},\label{eq:ideal-speech-estimation}
\end{equation} 
i.e., a matrix whose rows can be applied directly in \cref{eq:channel-estimation} to obtain the desired estimate of the log-magnitude of the \gls{rir} in the frequency domain.

%% file: sec-approach-minimum-phase.tex
\subsection{Minimum-phase Digital Filter Estimation}
The estimated channel log-magnitude $\widehat{\underbar H}(f) \approx \log \abs{ H(f) }$ is not sufficient to recover an estimate $\hat h(t)$ of the \gls{rir} in the time domain by using the inverse Fourier transform: This operation requires an estimate of the phase component $\angle H(f)$, which is not retrieved by the algorithm in \cite{Gaubitch:ChannelEstimation}.

We thus propose to design a causal digital filter $\widehat H(z)$ with a response as close as possible to $\widehat{\underbar H}(f)$, and retrieve the \gls{rir} $\hat h(t)$ by filtering a unit impulse with $\widehat H(z)$.

Let us denote with $H(e^{j\omega})\in\mathbb{C},\,-\pi<\omega\leq +\pi$ the channel response we want to obtain, and assume our desired digital filter to be defined by
\begin{equation}
    \widehat H(z) = \frac{\widehat B(z)}{\widehat A(z)},
\end{equation}
where
\begin{equation}
\begin{split}
    \widehat B(z) & = \hat b_0 + \hat b_1 z^{-1} + \ldots + b_{n_b} z^{-n_b}, \\
    \widehat A(z) & = 1 + \hat a_1 z^{-1} + \ldots + a_{n_a} z^{-n_a}. \\
\end{split}    
\end{equation}
The coefficients of the digital filter $\widehat H(z)$ can be obtained by minimizing the $l^2$-norm of the error
\begin{equation}
   J(\hat \theta) = \lnorm{ H(e^{j\omega}) - \widehat H(e^{j\omega})}
\end{equation}
with respect to the filter coefficients
\begin{equation}
   \hat \theta = [ \hat b_0, \hat b_1, \ldots, \hat b_{n_b}, \hat a_1, \hat a_2, \ldots, \hat a_{n_a}]
\end{equation}
by applying, e.g., Prony's method for filter design~\cite{Smith:2007:FilterDesign,Burrus:1970:FilterDesign}.

The highest stability and accuracy of filter design methods is achieved in the case of minimum phase filters, for which poles and zeros fall inside the unitary circle. The phase of these filters can be determined analytically  by recalling that the logarithm of their magnitude response is related to the phase response by means of the Hilbert transform $\mathcal H(\cdot)$~\cite{oppenheim}:
\begin{equation}
    \mathcal H \left(  \log \abs{H(f)} \right) = \log \abs{H(f)}   - j  \angle H(f). 
\end{equation}
The minimum-phase digital filter $\widehat H(z)$ should therefore be designed to minimize its $l^2$-norm with respect to the transfer function $\widehat H(f)$ defined by:
\begin{equation}
    \widehat H(f) = %
    \operatorname{exp}\left\lbrace \widehat{\underbar H}(f)\right\rbrace %
    \operatorname{exp}\left\lbrace -j \operatorname{Im}\left( \mathcal H \left( \widehat{\underbar H}(f) \right)\right)\right\rbrace,
\end{equation}
with $\operatorname{Im}(\cdot)$ denoting the imaginary part of a complex number and $\operatorname{exp}\lbrace\cdot\rbrace$ the exponential operator.

%% file: sec-approach-roomprint.tex
\subsection{Blind Roomprint Estimation}
A roomprint describing the \gls{rir} associated to the digital filter $\widehat H(z)$ can be retrieved according to the procedure outlined in \Cref{sec:background}.

The first step consists of recovering $\hat h(t)$, i.e., the estimated \gls{rir} in the time domain:
\begin{equation}
    \hat h(t) = \operatorname{digital-filter}\left( \delta(t), \widehat H(z) \right),
\end{equation}
where $\delta(t)$ is the unit impulse:
\begin{equation}
    \delta(t) =
    \begin{cases}
      1 & \text{ if $t=0$}, \\
      0 & \text{ if $t\neq0$}.
    \end{cases}
\end{equation}

$\hat h(t)$ is then filtered with fractional octave bandpass filters, to obtain band-limited signals
\begin{equation}
    \hat h^{(b)}(t) = \operatorname{bandpass-filter}( \hat h(t), f^{(b)}_l, f^{(b)}_u ),
\end{equation}
with $f^{(b)}_l$ and $f^{(b)}_u$ being the lower and upper bandedge frequencies of the $b$-th filter.

Finally, the estimated roomprint $\hat \psi$ can be computed by collecting several \glspl{rt60} estimated independently for each band using the Schroeder's integration method:
\begin{equation}\label{eq:roomprint-estimate}
    \hat \psi = \left[ \widehat \RT_{60}^{(1)}, \widehat \RT_{60}^{(2)}, \ldots, \widehat \RT_{60}^{(b)}, \ldots, \widehat \RT_{60}^{(\mathcal{B})} \right].
\end{equation}

%% file: sec-approach-svm.tex
\subsection{Closed-Set Classification}
The last step of our proposed approach for environment classification consists of a feature vector computation, and of the actual training of the classifier. To ease the reproducibility, we use the roomprint estimate $\hat \psi$ in \cref{eq:roomprint-estimate} as feature vector for the classification. 

The algorithm selected for the classification is a classic \gls{svm} with \gls{rbf} kernel, with hyperparameters $c$ and $\gamma$ determined by grid search on the set  $\mathcal{G}_c \times \mathcal{G}_{\gamma}$, where $\mathcal{G}_c = \lbrace  10^{-4}, \ldots, 10^{3}\rbrace$ and $\mathcal{G}_\gamma = \lbrace 10^{-4}, \ldots, 10^{3}\rbrace$.

%% file: sec-evaluation.tex
\section{Evaluation}\label{sec:evaluation}

\subsection{Datasets Involved}
The evaluation of our proposed approach for environment classification involved several datasets in conjunction. 

The \gls{svm} used for environment classification was evaluated on a dataset created by combining the ACE corpus~\cite{dataset:ace-challenge} and the LibriSpeech corpus~\cite{dataset:librispeech}. 
The ACE corpus consists of \glspl{rir} of seven different rooms, the general information of which is reported in \Cref{tab:ACErooms}.
For each room the impulse response was recorded in two positions, first in near-field conditions, and then in far-field conditions. The speech recordings drawn from the LibriSpeech corpus, which present no reverberation nor background noise, were first resampled to 16\,kHz and then convolved with the seven pairs of \glspl{rir} from the ACE corpus.

\begin{table}[hb]
    \centering
    \caption{Dimensions and \glspl{rt60} of rooms present in the ACE corpus for acoustic parameter estimation~\cite{dataset:ace-challenge}}
    \label{tab:ACErooms}
    \resizebox{\columnwidth}{!}{%
    \begin{tabular}{@{}llrrrrr@{}}
        \toprule
        \multirow{2}{*}{Label $l$}  &  \multirow{2}{*}{Name} & \multicolumn{3}{c}{Size (m)} & \multirow{2}{*}{Volume (m\textsuperscript{3})} & \multirow{2}{*}{$\RT_{60}$ (s)}\\ \cmidrule(lr){3-5}
         & & L & W & H &   &   \\
        \midrule
        1 & Office 1 & 4.8 & 3.3 & 3.0 & 47 & 0.34 \\
        2 & Office 2 & 5.1 & 3.2 & 2.9 & 48 & 0.39 \\
        3 & Meeting Room 1 & 6.6 & 4.7 & 3.0 & 92 & 0.44 \\
        4 & Meeting Room 2 & 10.3 & 9.2 & 2.6 & 250 & 0.37 \\
        5 & Lecture Room 1 & 6.9 & 9.7 & 3.0 & 200 & 0.64 \\
        6 & Lecture Room 2 & 13.4 & 9.2 & 2.9 & 360 & 1.25 \\
        7 & Building Lobby & 5.1 & 4.5 & 3.2 & 72 & 0.65 \\
        \bottomrule
    \end{tabular}}
\end{table}

The generated dataset consists of 200 noiseless reverberant recordings per \gls{rir} to be used for evaluation, with a total of 1400 recordings in near-field conditions as well as  1400 recordings in far-field conditions. The two sets were then split \emph{once}, retaining 80\% of the content for training and validation, and 20\% for testing. Each set can be uniquely identified as follows:
\begin{align*}
    X^\text{train}_\text{near}&: \text{Training examples in \emph{near-field} conditions}\\
    X^\text{test}_\text{near}&: \text{Test examples in \emph{near-field} conditions}\\
    X^\text{train}_\text{far}&: \text{Training examples in \emph{far-field} conditions}\\
    X^\text{test}_\text{far}&: \text{Test examples in \emph{far-field} conditions}
\end{align*}

To avoid any bias in the evaluation, we ensured an equal proportion of samples for each room in the training and testing set. Furthermore, no speech sample was convolved with two or more rooms, nor it appears in both the test and training set. 

The last dataset involved in the evaluation is the VCTK corpus~\cite{dataset:vctk}, which we used to train the \gls{gmm} involved in the channel estimation procedure. In particular, we used the whole content of VCTK to train a 1024 mixture \gls{gmm}, using 12 \glspl{mfcc} per frame, where each frame was processed with an Hanning window having 128\, ms of length and 50\% overlap, which in \cite{Gaubitch:ChannelEstimation} were found being optimal\footnote{Their optimality can also be confirmed by comparing different \gls{gmm} configurations by means of the \gls{bic}.}. The speakers present in the VCTK corpus are absent in the LibriSpeech dataset related to the \gls{svm} training and testing. Therefore, this choice for the \gls{gmm} should help in avoiding any evaluation bias related to the outcome of the environment classification. The sampling frequency was forced being equal to 16\,kHz, to ensure compatibility between the channel estimation and the classification stage.

\subsection{Bandwidth of the \texorpdfstring{\nicefrac{1}{B}}{1/B}-octave filters}

The first experiments to determine the quality of the environment classification aimed at determining the optimal bandwidth of the \nicefrac{1}{B}-octave fractional filters. 

We tested the system training the \gls{svm} on roomprints computed using \nicefrac{1}{3}-octave, \nicefrac{1}{4}-octave and \nicefrac{1}{8}-octave fractional filters, obtaining the performance reported in \Cref{tab:eval:filter-bandwidth}. All cases refer to \glspl{rt60} computed by linear interpolating $\RT_{20}$ values on the near-field dataset $X^\text{test}_\text{near}$.

The configuration with \nicefrac{1}{4}-octave bandwidth, corresponding to 26 bands ranging approximately from 100\,Hz to 8\,kHz, is the one performing the best, with an average accuracy of 89.6\%. The outcome confirms the finding in \cite{Roomprints:AES:2014}, despite the higher sampling frequency used in our experiments. 

\begin{table}[hb]
    \centering
    \caption{Impact of filter bandwidths}\label{tab:eval:filter-bandwidth}
    \begin{tabular}{@{}lrrr@{}}
        \toprule
        \multirow{2}{*}{Bandwidth} &  \multicolumn{3}{c}{Performance (\%)} \\ \cmidrule(l){2-4}
                     & Precision & Recall & Accuracy \\ \midrule
        \nicefrac{1}{3}-octave & 86.5 & 86.1 & 86.1\\
        \nicefrac{1}{4}-octave & 90.2 & 89.6 & 89.6\\
        \nicefrac{1}{8}-octave & 87.9 & 87.5 & 87.5 \\ \bottomrule
    \end{tabular}
\end{table}

\subsection{Interpolation parameter for RT60 estimation}

A second experiment addressed the interpolation parameter $\alpha$ involved in the estimation of the \glspl{rt60}, once again in relation to the near-field test set $X^\text{test}_\text{near}$. 

In particular, we set $\alpha$ in order to base the estimation of the reverberation time on the value of $\RT_{20}$ ($\alpha=3$), $\RT_{30}$ ($\alpha=2$), $\RT_{40}$ ($\alpha=1.5$), $\RT_{50}$ ($\alpha=1.2$) and $\RT_{60}$ itself ($\alpha=1$), obtaining the performance reported in \Cref{tab:eval:rt60alpha}. According to the previous experiments, the filter bandwidth was fixed and set equal to \nicefrac{1}{4}-octaves.

The configuration based on $\RT_{40}$ ($\alpha=1.5$) is the one performing best. Intuitively, lower values of $\alpha$ at first improve the performance of the classification, since the uncertainty is decreasing accordingly. Very low values of $\alpha$, however, cause numerical instability for bands that decrease too rapidly, since the Schroeder's integration method is disrupted by the floor noise level. This experiments confirms the tendency, and allows us to determine $\alpha=1.5$ as sweet spot for the evaluation.

\begin{table}[hb]
    \centering
    \caption{Impact of \gls{rt60} interpolation}\label{tab:eval:rt60alpha}
    \begin{tabular}{@{}lcrrr@{}}
        \toprule
        \multirow{2}{*}{$\RT_{60/\alpha}$} & \multirow{2}{*}{$\alpha$} &  \multicolumn{3}{c}{Performance (\%)} \\ \cmidrule(l){3-5}
                   &  & Precision & Recall & Accuracy \\ \midrule
        $\RT_{20}$ & 3.0 & 90.2 & 89.6 & 89.6\\
        $\RT_{30}$ & 2.0 & 90.9 & 90.7 & 90.7\\
        $\RT_{40}$ & 1.5 & 93.8 & 93.6 & 93.6 \\
        $\RT_{50}$ & 1.2 & 91.6 & 91.1 & 91.1 \\
        $\RT_{60}$ & 1.0  & 90.7 & 90.4 & 90.4 \\\bottomrule
    \end{tabular}
\end{table}

\subsection{Near-field and Far-field Comparison}

In the third and last experiment, we used the parameters selected for being the most effective -- namely a \nicefrac{1}{4}-octave filter bandwidth and the \gls{rt60} interpolation parameter $\alpha=1.5$ -- to evaluate the performance of the system in relation to near-field and far-field recording conditions. 

The experiments are summarised in \Cref{tab:eval:distance}, in which the last rows refer to the case of performing the evaluation by merging the near-field and far-field datasets together. The outcome of the evaluation, which we are going to comment in the following, met our expectations in terms of the predicted system behavior.

In near-field conditions for both training and test data, the system performs the best with a satisfying accuracy of 93.6\%, whereas in far-field conditions the accuracy drops to 89.6\%. The drop is probably due to the difficulty of the \gls{gmm} in estimating the ideal speech, due to the low ratio between the energy of the direct speech for the frame, and the one of the reverberant speech from previous ones. 

When training and testing conditions do not match, the accuracy drops significantly by more than 20\%. Once again, the drop could have been predicted beforehand: The distribution of training and test data are very different -- to the point that is debatable whether the room should be identified or not, considering that the \glspl{rir} might differ significantly -- and thus the \gls{svm} has severe problems in the classification task.

\begin{table}[b]
    \centering
    \caption{Impact of Near-field and Far-field conditions}\label{tab:eval:distance}
    \begin{tabular}{@{}lcrrr@{}}
        \toprule
        \multirow{2}[3]{*}{\splitcell[c]{Training\\Dataset}} & \multirow{2}[3]{*}{\splitcell[c]{Test\\Dataset}} &  \multicolumn{3}{c}{Performance (\%)} \\ \cmidrule(l){3-5}
                   &  & Precision & Recall & Accuracy \\ \midrule
        $X^\text{train}_\text{near}$ & $X^\text{test}_\text{near}$ & 93.8 & 93.6 & 93.6\\[0.25em]
        $X^\text{train}_\text{far}$ & $X^\text{test}_\text{far}$   & 90.2 & 89.6 & 89.6\\[0.25em]
        $X^\text{train}_\text{near}$ & $X^\text{test}_\text{far}$  & 71.1 & 70.0 & 70.0 \\[0.25em]
        $X^\text{train}_\text{far}$ & $X^\text{test}_\text{near}$  & 62.7 & 63.9 & 63.9 \\[0.25em]
        $X^\text{train}_\text{mixed}$ & $X^\text{test}_\text{mixed}$ & 84.8 & 84.6 & 84.6\\\bottomrule
    \end{tabular}
\end{table}

Lastly, the performance of the \gls{svm} in mixed conditions is better than for mismatching training and testing sets, but still inferior to the initial one with training and test set properly selected. Indeed, we can imagine that the \gls{svm} is fitting two (in principle disjoint) distributions per room, leading to a high error rate.

To summarize, this last experiment proves that the proposed algorithm for environment classification should be applied only in presence of prior information about the relative distance between the speakers and the recording device, in order to select the training content appropriately.

\subsection{Relation with state-of-the art results}

Existing state-of-the-art methods, as stated upfront in the introduction, can be applied in the presence of impulsive signals created e.g. by hand claps, which however rarely occur in relevant audio recordings~\cite{Roomprints:AES:2014,Roomprints:IEEE:2013,EnvClass:Patole:CAST,EnvClass:Malik:ICASSP}. 

A comparison might have been possible by applying an algorithm for semi- or fully- automatic selection of reverberation tails, and by processing these tails as if they were created by an impulsive sound or white noise rather than by speech. 

We decided not to perform such a comparison, however, to avoid incurring in any bias due to, e.g., faulty tail selection or mismatching statistical properties of the input audio signal. As absolute reference, the state-of-the-art roomprints proposal in~\cite{Roomprints:AES:2014} achieved a classification accuracy of 97.1\% on a self-curated dataset with 22 rooms, given noiseless impulse responses as input.

%% file: sec-outlook.tex
\section{Conclusions and Outlook}\label{sec:outlook}
To our knowledge, the algorithm proposed in this work represents the first attempt to address environment classification for the specific case of speech recordings: Previous methods addressed recordings created under laboratory conditions, exploiting decaying reverberation tails produced by impulsive signals, resulting in difficulties to obtain stable estimates of the reverberation parameters. In contrast, our approach is based on multi-band \gls{rt60} analysis of blind channel estimates, which does not require selection of the decaying reverberation tails, and is suitable for retrieving roomprint features whenever speech is dominant in a recording.

Under near-field and noiseless conditions, the algorithm is achieving an accuracy of about 93.6\% for recordings derived from the ACE corpus, dropping to about 89.6\% in the case of far-field conditions. The accuracy of the system drops significantly in mismatching conditions, thereby providing implicit evidence of the sensibility of the algorithm to changes with respect to the room impulse responses of speech source and recording device. 

For the future, we plan to determine whether noisy conditions are best addressed using the original derivation for channel estimation in \cite{Gaubitch:ChannelEstimation}, or rather by applying AI-based denoising as in \cite{Cuccovillo:2022:MAD,Giganti:2022:Euspico}. Furthermore, since the channel estimation procedure has been extensively used for performing microphone classification, as e.g. in~\cite{Cuccovillo:2022:MAD,Giganti:2022:Euspico} and references thereof, we also plan to investigate the influence of the microphone on environment classification, and to determine whether the two contributions to the channel can be decoupled effectively or not.